\documentclass[aps, prl, onecolumn, superscriptaddress]{revtex4-2}

\usepackage{graphicx,color}
\usepackage{dcolumn}
\usepackage{bm}
\usepackage{amssymb}
\usepackage{textcomp}
\usepackage[section]{placeins}
\usepackage{float}
\usepackage{epsfig,amssymb,amsmath,amsthm,amsfonts,amsbsy,mathrsfs}
\usepackage{eqnarray}
\usepackage{ifthen}
\usepackage{multirow}
\usepackage{array}
\usepackage[displaymath]{lineno}
\usepackage{subfigure}
\usepackage[capitalise]{cleveref}
\usepackage{comment}
\usepackage{natbib}

\NewDocumentCommand{\codeword}{v}{ \texttt{#1} }
\begin{document}

\title{Analytical Calculation of Viscosity in Rouse Networks Below Gelation Transition}
\author{Bohan Lyu} 
    \affiliation{Peking-Tsinghua Center for Life Sciences, Peking University, Beijing, China}
\author{Jie Lin}
\affiliation{Peking-Tsinghua Center for Life Sciences, Peking University, Beijing, China}
\affiliation{Center for Quantitative Biology, Peking University, Beijing, China}
\date{\today}

\begin{abstract}
This work establishes an exact relationship between the zero-shear viscosity and the radius of gyration for generalized Rouse model with arbitrary network configurations. Building on this fundamental relationship, we develop a random graph approach to derive analytical expressions for network viscosity in mean-field connected systems through coarse-grained graph viscosity relations. The theory fully characterizes the subcritical regime, determining precise asymptotic behavior in both dilute and pre-gelation limits.
\end{abstract}

\maketitle

\subsection{A. Introduction}
The Rouse model predicts the viscosity of unentangled linear polymer chains in dilute solutions through bead-spring dynamics \cite{Rouse1953}. Subsequent generalizations have extended the Rouse framework to polymer networks, where bead connectivity topology dictates both conformational statistics and viscoelastic properties \cite{Yang1998, cohen_direct_2024}. However, deriving rigorous analytical solutions for viscosity in randomly crosslinked systems remains a fundamental challenge.

The radius of gyration ($ R_g $) serves as a pivotal experimental observable for quantifying chain compactness via scattering techniques \cite{Higgins1994}. Theoretical foundations for $\langle R_g^2 \rangle$ have been established across canonical polymer configurations, including ideal chains and regular networks \cite{Flory1969}. $ R_g $ has been also shown to govern steady-shear viscosity through microstructure-rheology coupling \cite{uneyama_radius_2025}.

This work developed a comprehensive theoretical framework for generalized Rouse network rheology, where we discover a simple proportionality between zero-shear viscosity and the equilibrium radius of gyration for the network with arbitrary configurations. We establish analytical solutions that quantify viscosity in randomly crosslinked networks in pre-gelation regimes, revealing how connectivity patterns dictate macroscopic flow behavior through rigorous mathematical formalism.

\subsection{B. Rouse Network}
We consider a connected polymer network of $N$ beads with arbitrary configurations. In the original Rouse model, the network is a linear chain\cite{Rouse1953}. Here, we consider a generalized Rouse network with an arbitrary configuration\cite{yang_graph_1998,cohen_direct_2024}. The position of bead $i$ in three dimensions, $\bm{R}_i$, follows an overdamped dynamics
\begin{equation}
	\zeta \dot{\bm{R}}_i = - k L_{ij} \bm{R}_j + \bm{\xi}_i
	\label{eq:dynamics}
\end{equation}
Here, $\zeta$ is the friction coefficient, $k$ is the spring constant, $\bm{\xi}_i$ is the random force due to thermal fluctuation, which satisfies the fluctuation-dissipation theorem,
\begin{equation}
	\langle \xi_{i,\alpha}(t) \xi_{j,\beta}(t') \rangle = 2\zeta k_B T \delta_{ij}\delta_{\alpha\beta} \delta(t-t').
	\label{eq:fdt}
\end{equation}
Here $\delta_{ij}$ and $\delta_{\alpha\beta}$ are Kronecker delta functions, and $\delta(t-t')$ is Dirac delta function. The Greek letters represent the directions in the Cartesian coordinate. $L_{ij}$ is the $N\times N$ graph Laplacian matrix that describes the connectivity between beads,	 and its non-zero eigenvalues are $\lambda_p \geq 0$ with orthonormal eigenvectors $\bm{u}_p$.

We project the $N \times 3$ position vector $\bm{R}$ onto the eigenmodes such that $\bm{c}_p = \bm{u}_p^T \bm{R}$, where $\bm c_p$ is a $1 \times 3$ vector which satisfies
\begin{equation}
    \zeta \dot{\bm{c}}_p = -k \lambda_p \bm{c}_p + \bm{f}_p,
    \label{eq:modal}
\end{equation}
where $\bm{f}_p = \bm{u}_p^T \bm{\xi}_i$, which satisfies
\begin{equation}
    \langle f_{p,\alpha}(t) f_{q,\beta}(t') \rangle = 2\zeta k_B T \delta_{pq}\delta_{\alpha\beta} \delta(t-t')
\end{equation}

\subsection{C. Stress Relaxation Dynamics}
We compute the macroscopic stress tensor $\sigma_{xy}$ from the microscopic forces and bead positions
\begin{equation}
    \sigma_{xy} = \frac{\sum_i F_{i,x}y_i}{V}= \frac{k}{V} \langle \bm{y} | \bm{L} | \bm{x} \rangle
    = \frac{k}{V} \sum_{p} \lambda_p c_{px} c_{py}
    \label{eq:stress}
\end{equation}
where $F_{i,x}$ is the $x$ component of the total force applied on bead $i$, $c_{px}$, $c_{py}$ are the $x$ and $y$ components of $\bm c_p$. Note that the summations over $p$ does not include zero eigenvalues. Given a step strain $\gamma$ at $t=0$ along $x$-direction, we have
\begin{align}
    y_i(t=0^+) &= y_i(t=0^-),\\
    x_i(t=0^+) &= x_i(t=0^-) + \gamma y_i(t=0^-),
\end{align}
which means that
\begin{align}
     c_{py}(t=0^+) &=c_{py}(t=0^-), \\
   c_{px}(t=0^+) &=c_{px}(t=0^-) + \gamma  c_{py}(t=0^-)
\end{align}
The correlation at time zero between $c_{px}$ and $c_{py}$ can be obtained from the equipartition theorem:
\begin{equation}
    \langle c_{px}(0^+) c_{py}(0^+) \rangle = \gamma \langle c_{py}^2 \rangle_{\text{eq}} 
    = \gamma \frac{k_B T}{k \lambda_p}
    \label{eq:initial_corr}
\end{equation}

The time evolution of the correlation function follows from \eqref{eq:modal}:
\begin{align}
    \frac{d}{dt} \langle c_{px} c_{py} \rangle 
    &= \langle \dot{c}_{px} c_{py} \rangle + \langle c_{px} \dot{c}_{py} \rangle \\
    &= -\frac{2k\lambda_p}{\zeta} \langle c_{px} c_{py} \rangle
\end{align}
Solving with the initial condition \eqref{eq:initial_corr}, we obtain
\begin{equation}
    \langle c_{px}(t) c_{py}(t) \rangle = \frac{\gamma k_B T}{k \lambda_p} e^{-t / \tau_p},
    \label{eq:corr_decay}
\end{equation}
where $\tau_p = \frac{\zeta}{2k \lambda_p}$. Substituting into \eqref{eq:stress} gives the stress relaxation, we obtain
\begin{equation}
    \langle \sigma_{xy}(t) \rangle = \frac{\gamma k_B T}{V} \sum_{p} e^{-t / \tau_p}.
    \label{eq:stress_relax}
\end{equation}
The zero-shear viscosity $\eta$ is the time integral of stress relaxation:
\begin{equation}
    \eta = \frac{1}{\gamma} \int_0^\infty \langle \sigma_{xy}(t) \rangle  dt = \frac{\zeta k_B T}{2k V} \sum_{p} \frac{1}{\lambda_p}
    \label{eq:viscosity}
\end{equation}
We define the dimensionless parameter $\tilde{\eta}$ as:
\begin{equation}
\tilde{\eta} = \frac{2k V}{\zeta k_B T}\eta = \sum_{\lambda \neq 0} \lambda^{-1} 
\label{eq:structural-sum}
\end{equation}
Throughout our following analysis, \textbf{``viscosity''} and the dimensionless parameter $\mathbf{\tilde{\eta}}$ will be used interchangeably.

\subsection{D. Relationship between Radius of Gyration and Viscosity}
For a general polymer network, the radius of gyration can be computed as \cite{Rubinstein2003}
\begin{equation}
	\langle R_g^2\rangle = \frac{1}{2N^2}\sum_{i,j} (\bm{R}_i-\bm{R}_j)^2.
\end{equation}
We write $x_i-x_j = (\bm e_i- \bm e_j)^T \sum_p (\bm c_{px} \bm u_p)$, where $\bm e_i$ is a $N\times 1$ vector with its $i$ element equal to one and all other elements equal to zero. Since the eigenvector corresponding to the zero eigenvalue represents a rigid-body translation mode whose eigenvector components are all identical, we have $(\bm e_i- \bm e_j)^T\bm u_p=0$ for $\lambda_p=0$; therefore, the summation only includes nonzero eigenvalues. We obtain
\begin{equation}
\begin{aligned}
	\sum_{i,j}\langle (x_i-x_j)^2\rangle &= \sum_{i,j} \langle (\bm e_i- \bm e_j)^T \sum_p (c_{px} \bm u_p) \sum_q (c_{qx} \bm u_q)^T (\bm e_i- \bm e_j)\rangle\\
	&=\langle \sum_{p,q} c_{px}c_{qx} \sum_{i,j} (u_{pi}-u_{pj})(u_{qi}-u_{qj})   \rangle\\
	&=\sum_p \langle c_{px}^2\rangle \sum_{i,j} (u_{pi}^2+u_{pj}^2-2u_{pi}u_{pj})\\
	&=\sum_p \frac{2k_BT}{k\lambda_p } [N \sum_i u_{pi}^2 - (\sum_i u_{pi})^2]\\
	&=\sum_p  \frac{2Nk_BT}{k\lambda_p }.
\end{aligned}
\end{equation}
Here, we have used the orthonormal condition of $\sum_i u_{pi}^2 =1$. Because the null space of the $L_{ij}$ matrix is spanned by a $N\times 1$ vector with all elements equal to one, we also have $\sum_i u_{pi}=0$. Since the three directions are equivalent, we immediately obtain

\begin{equation}
    \langle R_g^2 \rangle = \frac{3k_B T}{N k} \sum_{p} \frac{1}{\lambda_p}
    \label{eq:rg}
\end{equation}
Combining \eqref{eq:viscosity} and \eqref{eq:rg}, we obtain a key result:
\begin{equation}
\eta = \frac{N \zeta}{6V} \langle R_g^2 \rangle 
    \label{Rg_eta}
\end{equation}

For tree-like Rouse networks, we apply Kramers' theorem to determine the mean square radius of gyration $\langle R_g^2 \rangle$ \cite{Rubinstein2003}. Defining $\pi_e$ as the product of the component sizes formed upon removal of bond $e$, the radius of gyration follows:

\begin{equation}
\langle R_g^2 \rangle = \frac{b^2}{N^2} \sum_e \pi_e
\label{Kramers_Rouse}
\end{equation}

Substituting \eqref{Kramers_Rouse} into the viscosity relation \eqref{Rg_eta}, and noting that the equipartition theorem for a harmonic spring in three dimensions ($\frac{1}{2}k b^2 = \frac{3}{2}k_B T$) implies $b^2 = 3k_B T / k$, we obtain:

\begin{equation}
\eta = \frac{\zeta k_B T}{2kV N} \sum_e \pi_e
\label{realeta_pi}
\end{equation}
Substituting \eqref{realeta_pi} into \eqref{eq:structural-sum} yields the normalized viscosity for tree-like networks:

\begin{equation}
\tilde{\eta} = \frac{1}{N} \sum_e \pi_e
\label{eq:eta-pi}
\end{equation}

For systems with multiple connected tree-like components, the viscosity is given by:
\begin{equation}
\tilde{\eta} = \sum_{\alpha} \frac{1}{N_\alpha} \sum_{e \in G_\alpha} \pi_e^{(G_\alpha)}
\label{eta_components}
\end{equation}
where $\alpha$ indexes connected components, $N_\alpha$ is the size of component $\alpha$, and $\pi_e^{(G_\alpha)}$ is computed within that component.

\subsection{E. Calculate the Viscosity of a Network from Its Coarse-Grained Representation}

We consider a connected tree-like network of $N_p$ polymer chains, each containing $N_m$ beads. The network topology is defined by a coarse-grained tree graph $G$ where each node represents a polymer chain, and edges represent crosslinks between chains. For each edge in $G$, crosslinks are formed by randomly selecting one bead from each connected polymer chain.

The total viscosity $\tilde{\eta}$ follows from the bond representation (Eq. \ref{eq:eta-pi}):
\begin{equation}
N_p N_m \tilde{\eta} = \sum_e \pi_e = \sum_{e_c} \pi_{e_c} + \sum_{e_b} \pi_{e_b}
\label{eq:total-viscosity}
\end{equation}
where $e_c$ are crosslinks between polymers and $e_b$ denotes backbone bonds within each polymer chain. These contributions will be calculated separately in the following subsections.

\textbf{1. Crosslink Contribution}

Each crosslink corresponds to an edge $e_c'$ in the coarse-grained graph $G$. The contribution is:
\begin{equation}
\pi_{e_c} = N_m^2 \pi_{e_c'}^{G}
\label{eq:crosslink-pi}
\end{equation}
Summing over all crosslinks yields:
\begin{equation}
\sum_{e_c} \pi_{e_c} = \sum_{e_c'} N_m^2 \pi_{e_c'}^{G} = N_m^2 (N_p \tilde{\eta}_{G})
\label{eq:crosslink-sum}
\end{equation}
where $\tilde{\eta}_{G}$ is the viscosity of $G$ according to Eq. \eqref{eq:eta-pi}.

\textbf{2. Backbone Bond Contribution}

To systematically account for paths traversing backbone bonds, we establish three key elements of notation: first, the index $s$ ranges over all polymer chains ($s = 1,2,\dots,N_p$) and $\bar{s}$ denotes the complement comprising all beads not in chain $s$; second, the indicator function $(ij,e_b)$ equals 1 if the path between nodes $i$ and $j$ traverses bond $e_b$ (0 otherwise); third, node pairs are classified into three categories: pairs $(i,j)\in s\times s$ within the same polymer $s$, pairs $(i,j)\in s\times \bar{s}$ with one node in $s$ and one outside, and pairs $(i,j)\in \bar{s}\times \bar{s}$ where both nodes lie outside $s$ but paths traverse bonds within $s$.

The backbone bond contribution then decomposes naturally as:
\begin{equation}
\begin{aligned}
\sum_{e_b} \pi_{e_b} = & \sum_s \sum_{\substack{(i,j)\in s\times s \\ e_b\in s}} (ij,e_b) 
+ \sum_s \sum_{\substack{(i,j)\in s\times \bar{s} \\ e_b\in s}} (ij,e_b) 
+ \sum_s \sum_{\substack{(i,j)\in \bar{s}\times \bar{s} \\ e_b\in s}} (ij,e_b)
\label{eq:backbone-decomp}
\end{aligned}
\end{equation}

These three components will be calculated separately in the following.

\textbf{2.1. $s\times s$ Component}

This contribution represents all paths within a single polymer chain. For any chain $s$, the sum over pairs $(i,j)$ and paths traversing backbone bonds $e_b$ is equivalent to:
\begin{equation}
\sum_{\substack{(i,j)\in s\times s \\ e_b\in s}} (ij,e_b) 
= \frac{1}{2} \sum_i\sum_j \lvert i - j \rvert
= N_m \tilde{\eta}_p
\end{equation}
where 
\begin{align}
\tilde{\eta}_p = \frac{1}{N_m}\sum_{i<j} |i-j| = \frac{N_m^2 - 1}{6}
\label{eq:etap}
\end{align}
is the viscosity of an isolated linear chain with $N_m$ beads, as defined in \eqref{eq:eta-resistance}.

Summing over all $N_p$ chains yields:
\begin{equation}
\sum_s \sum_{\substack{(i,j)\in s\times s \\ e_b\in s}} (ij,e_b) = N_p N_m \tilde{\eta}_p
\label{eq:intra-term}
\end{equation}

Additionally, we define the average path length $\bar{d}$ within a linear chain as the average distance between any two beads (including same bead pairs):
\begin{equation}
\bar{d} = \frac{1}{N_m^2} \sum_i\sum_j \lvert i - j \rvert 
\label{eq:d_bar_chain}
\end{equation}
which relates to the viscosity by:
\begin{equation}
\bar{d} = \frac{2\tilde{\eta}_p}{N_m}
\label{eq:dbar}
\end{equation}

\textbf{2.2. $s\times \bar{s}$ Component}

This component quantifies paths between nodes in polymer chain $s$ and nodes in other polymer chains ($\bar{s}$). There are $N_m$ possible nodes in $s$ and $(N_p-1)N_m$ nodes in other polymer chains, giving $N_m^2 (N_p-1)$ such node pairs. Each path traversing $s$ has an average path length $\bar{d}$. Multiplying the number of pairs by this average path and substitute Eq.\eqref{eq:dbar} gives the per-chain contribution:

\begin{equation}
\sum_{\substack{(i,j)\in s\times \bar{s} \\ e_b\in s}} (ij,e_b) = N_m^2(N_p-1) \bar{d} = 2\tilde{\eta}_p N_m(N_p-1)
\end{equation}

Summing over all $N_p$ chains yields the total contribution:

\begin{equation}
\sum_s \sum_{\substack{(i,j)\in s\times \bar{s} \\ e_b\in s}} (ij,e_b) = 2 \tilde{\eta}_p N_m N_p (N_p - 1)
\label{eq:inter-term}
\end{equation}

\textbf{2.3. $\bar{s}\times \bar{s}$ Component}

This component accounts for paths between nodes in different branches that traverse bonds $e_b$ within polymer $s$. The paths pass through an average of $\bar{d}$ bonds within $s$. 

To compute this component, we define a branch size function $[s, l]$ as follows: for adjacent polymers $s$ and $l$ connected by a crosslink, $[s, l]$ is the number of polymer chains of the branch containing $s$ when the crosslink is removed; otherwise $[s,l]=0$.

By definition, the branch size function satisfies:
\begin{equation}
\begin{aligned}
[s, l] + [l, s] &= N_p \quad (s \neq l) \\
\sum_{l} [l, s] &= N_p - 1
\label{eq:sl_sum}
\end{aligned} 
\end{equation}

For each pair of adjacent polymers $l_1 \neq l_2$ connected to $s$, the number of path pairs $(i,j)$ is determined by the size of their respective branches. There are $N_m [l_1, s]$ nodes in branch $l_1$ and $N_m [l_2, s]$ nodes in branch $l_2$, yielding $(N_m [l_1, s])(N_m [l_2, s])$ possible node pairs. Using $\bar{d}$ from \eqref{eq:dbar} as the average number of bonds traversed within $s$, we obtain:
\begin{equation}
\begin{aligned}
\sum_s \sum_{\substack{(i,j)\in \bar{s}\times \bar{s} \\ e_b\in s}} (ij,e_b) &= \bar{d} \sum_s \sum_{l_1 < l_2} (N_m [l_1, s])(N_m [l_2, s])\\
&= 2\tilde{\eta}_p N_m\sum_s \sum_{l_1 < l_2} [l_1, s][l_2, s]
\label{sbarsbar}
\end{aligned}
\end{equation}

We compute the double summation using the branch size properties and viscosity definitions:
\begin{equation}
\begin{aligned}
\sum_s \sum_{l_1 < l_2} [l_1, s][l_2, s] &= \sum_s \frac{1}{2} \sum_{l_1} \sum_{l_2 \neq l_1} [l_1, s][l_2, s] \\
&= \frac{1}{2} \sum_s \sum_{l_1} [l_1, s] \left(\sum_{l_2 \neq l_1} [l_2, s]\right) \\
&= \frac{1}{2} \sum_s \sum_{l_1} [l_1, s] \left( \sum_{l_2} [l_2, s] - [l_1, s]\right) \\
&= \frac{1}{2} \sum_s \sum_{l_1} [l_1, s] \left((N_p - 1) - (N_p - [s,l_1])\right) 
\quad \text{(using \eqref{eq:sl_sum})} \\
&= \frac{1}{2} \sum_s \sum_{l_1} [l_1, s] \left( [s, l_1] - 1 \right) \\
&= \frac{1}{2} \sum_s \sum_{l_1} [l_1, s][s, l_1] - \frac{1}{2} \sum_s (N_p - 1) \\
&= \frac{1}{2} \sum_s \sum_{l_1} [l_1, s][s, l_1] - \frac{N_p(N_p - 1)}{2} \\
&= \tilde{\eta}_{G} N_p- \frac{N_p(N_p - 1)}{2}
\label{doublesum}
\end{aligned}
\end{equation}

In the last step, we used the viscosity formula \eqref{eq:eta-pi} for the coarse-grained graph $G$. Substituting \eqref{doublesum} into \eqref{sbarsbar} yields the final form:
\begin{equation}
\sum_s \sum_{\substack{(i,j)\in \bar{s}\times \bar{s} \\ e_b\in s}} (ij,e_b) = \tilde{\eta}_p N_p N_m \left( 2\tilde{\eta}_{G} - (N_p - 1) \right)
\label{eq:external-term}
\end{equation}

Substituting \eqref{eq:crosslink-sum}, \eqref{eq:intra-term}, \eqref{eq:inter-term}, \eqref{eq:external-term} and \eqref{eq:etap} into \eqref{eq:total-viscosity}, we finally get the viscosity of a tree-like network from its coarse-grained graph viscosity:
\begin{equation}
\begin{aligned}
\tilde{\eta} &= N_p\tilde{\eta}_p + (2\tilde{\eta}_p+N_m)\tilde{\eta}_{G} \\
&= \frac{N_p(N_m^2 - 1)}{6} + \frac{N_m^2 + 3N_m - 1}{3}\tilde{\eta}_{G}
\label{eq:full-equation}
\end{aligned}
\end{equation}

The viscosity increment $\Delta\tilde{\eta} = \tilde{\eta} - N_p \tilde{\eta}_p$ quantifies the added viscosity from crosslinking beyond isolated chains. This increment obeys the linear relation:
\begin{equation}
\Delta\tilde{\eta} = \frac{N_m^2 + 3N_m - 1}{3} \tilde{\eta}_{G}
\label{eq:viscosity-increment}
\end{equation}

Due to the the additivity of viscosity, Eq.\eqref{eq:viscosity-increment} also holds for tree-like networks with multiple components, where $\tilde{\eta}_{G}$ represents the viscosity of the entire coarse-grained network's graph Laplacian, which sums the contributions from all connected components.

\subsection{F. Viscosity of a Random Graph in the Subcritical Regime}

For a random graph $G(N_p, c)$ with $N_p \to \infty$ nodes and mean degree $c < 1$, the graph consists entirely of tree-like connected components. We derive the branch size distribution and viscosity based on the following foundations\cite{bollobas_random_2001}.

Consider a branch selection process in the graph: randomly select an edge, cut it, and randomly select one of the two resulting branch. We define $P(n)$ as the probability distribution of the branch size (number of nodes), with corresponding generating function $G(x) = \sum_{n=1}^{\infty} P(n)x^n$.

For the root node selected by branch selecting, the residual degree (number of remaining connections) exhibits a distribution derived as:
\begin{equation}
\begin{aligned}
P(\text{residual degree} = k) &= \frac{P(\text{degree} = k+1)(k+1)}{\sum_{k'}P(\text{degree} = k'+1)(k'+1)} \\
&= \frac{[e^{-c}c^{k+1}/(k+1)!](k+1)}{c}\\&=\frac{e^{-c} c^k}{k!}
\end{aligned}
\end{equation}
since the node degree distribution $P(\text{degree} = k)$ for randomly selected nodes is Poisson with mean $c$.

As the result, the generating function satisfies the self-consistency equation that encodes the tree-like structure:
\begin{equation}
\begin{aligned}
G(x) &= x \sum_{k=0}^{\infty} P(\text{residual degree}=k) [G(x)]^k \\
&= x e^{c(G(x) - 1)}
\label{Gx}
\end{aligned}
\end{equation}

This equation expresses the probabilistic structure of a branch: the $x$ factor represents the root node of the branch, $P(k)$ is the Poisson residual degree distribution of this root, and $[G(x)]^k$ accounts for $k$ independent subtrees attached to the root. This recursive form emerges because any branch can be viewed as a root node connected to a random number of stochastically identical branches, each described by the same generating function $G(x)$.

Applying the Lagrange inversion theorem to solve $x = G e^{c(1-G)}$, we get:
\begin{equation}
[x^n] G(x) = \frac{1}{n} [G^{-1}] \left( G e^{c(1-G)} \right)^{-n}
\label{eq:lagrange-formula}
\end{equation}
where $[x^n]$ and $[G^{-1}]$ denotes the coefficient extraction operator for the $x^n$ and $G^{-1}$ term in the series expansion.

Noticing $[x^n]G(x)=P(n)$ by definition, Eq.\eqref{eq:lagrange-formula} gives the branch size distribution:
\begin{equation}
P(n) = \frac{e^{-n c} (n c)^{n-1}}{n!}
\label{eq:branch-dist}
\end{equation}

The viscosity of the random graph can be calculated by Eq.\eqref{eta_components}:
\begin{equation}
\tilde{\eta}_G = \frac{N_p c}{2} \sum_{n_1=1}^{\infty} \sum_{n_2=1}^{\infty} P(n_1) P(n_2) \frac{n_1 n_2}{n_1 + n_2}
\label{eq:expected-viscosity}
\end{equation}
where we used the fact that there are $\frac{1}{2} N_p c$ edges in total.

Substituting \eqref{eq:branch-dist} into \eqref{eq:expected-viscosity}, we get the viscosity of the random graph:
\begin{equation}
\tilde{\eta}_G = N_p \sum_{n_1,n_2=1}^{\infty} 
\frac{e^{-c(n_1+n_2)} c^{n_1+n_2-1}n_1^{n_1} n_2^{n_2}}{2(n_1 + n_2) n_1! n_2!}
\label{eq:substituted-viscosity}
\end{equation}

If we define a scaling function $f(c)$ as:
\begin{equation}
f(c) := \sum_{n_1,n_2=1}^{\infty} 
\frac{e^{-c(n_1+n_2)} c^{n_1+n_2-1}n_1^{n_1} n_2^{n_2}}{2(n_1 + n_2) n_1! n_2!}
\label{eq:scaling-function}
\end{equation}
then we have:
\begin{equation}
\tilde{\eta}_G = N_p f(c)
\label{eq:substituted-viscosity}
\end{equation}

\subsection{G. Mean-Field Viscosity for Randomly Connected Polymer Chains}

Building on the structural relationship for network viscosity (Eq. \ref{eq:full-equation}) and the random graph viscosity (Eq. \ref{eq:substituted-viscosity}), we obtain the viscosity for a mean-field crosslinked Rouse network below the gelation transition :

\begin{equation}
\begin{aligned}
\tilde{\eta} &= N_p\tilde{\eta}_p + \frac{N_m^2 + 3N_m - 1}{3}N_pf(c) \\
&=  \frac{N_p(N_m^2 - 1)}{6} + \frac{N_p(N_m^2 + 3N_m - 1)}{6} \sum_{n_1,n_2=1}^{\infty} 
\frac{e^{-c(n_1+n_2)} c^{n_1+n_2-1}n_1^{n_1} n_2^{n_2}}{(n_1 + n_2) n_1! n_2!}
\end{aligned}
\end{equation}

The viscosity increment due to crosslinking is:
\begin{equation}
\Delta\tilde{\eta} = \frac{N_m^2 + 3N_m - 1}{3} N_p f(c)
\end{equation}

\subsection{H. Asymptotic Behavior of $f(c)$}

To compute the viscosity contribution for clusters of size $N$ in \eqref{eq:scaling-function}, we define:
\begin{equation}
f_N(c) = \sum_{\substack{n_1+n_2=N \\ n_1,n_2 \geq 1}} \frac{e^{-cN} c^{N-1} n_1^{n_1} n_2^{n_2}}{2N  n_1! n_2!}.
\label{eq:fNc}
\end{equation}

In the weakly crosslinked limit ($c \ll 1$), the scaling function $f(c)$ is dominated by the smallest cluster size ($N=2$ with $n_1=n_2=1$):
\begin{equation}
f(c) \approx f_2(c) = \frac{e^{-2c} c}{4} \approx \frac{c}{4}
\end{equation}
This linear dependence $f(c) \sim c$ reflects the viscosity enhancement due to crosslinking in the dilute crosslink regime.

Applying Stirling's approximation $n^n/n! \sim e^n/\sqrt{2\pi n}$ to \eqref{eq:fNc} yields:
\begin{equation}
\begin{aligned}
f_N(c) 
&\approx \frac{e^{-cN} c^{N-1}}{2N} \sum_{n_1=1}^{N-1} \frac{e^N}{2\pi \sqrt{n_1 n_2}} \\
&= \frac{e^{(1 - c)N} c^{N-1}}{4\pi N} \sum_{n_1=1}^{N-1} \frac{1}{\sqrt{n_1 (N - n_1)}} \\
&\approx \frac{e^{(1 - c)N} c^{N-1}}{4\pi N} \int_0^1 \frac{dx}{\sqrt{x(1-x)}} \\
&= \frac{e^{(1-c+\ln{c})N}}{4cN} 
\end{aligned}
\end{equation}

Near the gelation point ($c = 1 - \epsilon$ with $0 < \epsilon \ll 1$), the expansion $1 - c + \ln c \approx -\epsilon^2/2$ gives:
\begin{equation}
f_N(c) \approx \frac{e^{-\epsilon^2 N /2}}{4N}
\end{equation}
This describes an inverse proportionality to $N$ with an exponential cutoff at the characteristic cluster size $N^* \sim \epsilon^{-2}$. 

Using the Taylor expansion $\sum_{N=1}^{\infty} \frac{x^N}{N} = -\ln(1 - x)$ with $x = e^{-\epsilon^2/2}$, and noting the sum starts from $N=2$, the total viscosity scales as:
\begin{equation}
\begin{aligned}
f(c) &= \sum_{N=2}^{\infty} f_N(c) \\
&\approx \sum_{N=2}^{\infty} \frac{e^{-\epsilon^2 N /2}}{4N}\\
&\approx -\frac{1}{4}\ln\left(1 - e^{-\epsilon^2/2}\right) - \frac{e^{-\epsilon^2/2}}{4} \\
&\sim -\frac{1}{2}\ln\epsilon \quad \text{as} \quad \epsilon \to 0^+
\end{aligned}
\end{equation}

\subsection{I. Summary}
We establish a universal relation between zero-shear viscosity $\eta$ and equilibrium mean square radius of gyration $\langle R_g^2 \rangle$ for arbitrary Gaussian networks: $\eta = \frac{N\zeta}{6V} \langle R_g^2 \rangle$, enabling viscosity prediction directly from static conformation statistics. Building upon this fundamental correspondence, we derive exact analytical expressions for viscosity in randomly crosslinked polymer networks below the gelation transition. The theory reveals a linear proportionality between viscosity increment $\Delta\tilde{\eta}$ and coarse-grained graph viscosity, expressed as $\Delta\tilde{\eta} = \frac{N_m^2 + 3N_m - 1}{3} N_p f(c)$, where $N_m$ is chain size, $N_p$ is polymer count, $c < 1$ is mean crosslinking degree, and $f(c)$ is an analytic scaling function derived from random graph theory. This provides a complete mean-field solution for pre-gelation rheology.

\bibliography{MF}
\end{document}